# Smart fingertip sensor for food quality control: fruit maturity assessment with a magnetic device

Maria Carvalho[1,2,*], Pedro Ribeiro[1,2], Verónica Romão[1] Susana Cardoso[1,2],

[1]Instituto de Engenharia de Sistemas e Computadores – Microsistemas e Nanotecnologias (INESC MN), 1000-029 Lisbon, Portugal, mcarvalho@inesc-mn.pt; pribeiro@inesc-mn.pt; vromao@inesc-mn.pt; scardoso@inesc-mn.pt

[2]Instituto Superior Técnico, Universidade de Lisboa, Av. Rovisco Pais 1, 1049-001 Lisbon, Portugal

*Corresponding author at: Instituto de Engenharia de Sistemas e Computadores – Microsistemas e Nanotecnologias INESC-MN, Rua Alves Redol 9, 1000-029 Lisbon, Portugal.
E-mail address: mcarvalho@inesc-mn.pt

**ABSTRACT**

Automated technologies for quality inspection of fruits have attracted great interest in the food industry. The development of nondestructive mechanisms to assess the quality of individual fruit prior to sale may lead to an increase in overall product quality, value, and consequently, producer competitiveness. However, the existing methods have limitations.
Herein, a texture sensor based on highly sensitive hair-like cilia receptors, to allow a quick quality evaluation of fruit is proposed. The texture sensor consists of up to 100 magnetized nanocomposite cilia attached to a chip with magnetoresistive sensors in a full Wheatstone bridge architecture. In this paper we demonstrate the use of ciliary sensors in scanning fruits (blueberries and strawberries) in different maturation stages. The contact of the cilia with the fruit skin provided qualitative information about its texture in terms of ripeness stage. Less mature fruits exhibited, on average, a highest peak voltage of 0.14 mV for blueberries and 0.12 mV for strawberries, while overripe fruits exhibited 0.58 mV and 0.56 mV, respectively. The results were confirmed by sensorial assessment of the fruit freshness, and therefore attesting the application potential of the sensing technology for fruit quality control.

## 1. Introduction

There is a growing concern about diet and its influence on health. Fruits are among the most healthy and nutritive foods, namely in terms of vitamins and minerals, which are known to help in the prevention of various diseases [1]. Fruit consumption has increased, more than doubled in the past three decades [2], together with the demand for high quality products. Accordingly, the reputation of producers, and consequently their position in the market, is based on product quality, which makes quality controls essential [3]. Developing automated technologies capable of inspecting the quality of every piece of fruit prior to sale will increase product value and producer competitiveness.

Several nondestructive techniques have been reported for measuring quality attributes of fruit [4]. Techniques for external attributes such as color, size and absence of external defects are used on commercial sorting lines, mostly machine vision-based technologies. However, they present constraints as different fruit skin colors and defects as well as lighting conditions affect monitoring reliability [3]. For internal quality attributes, most techniques present a slow measurement speed, high cost and are specific for laboratory use, thus non-applicable for industrial sorting lines [5]. The success of these techniques depends on how closely they mimic the perceptions of humans, therefore future sensors should preferably be based on biomimetic principles [4].

Humans tend to rely on their sense of touch to grade fruits according to their ripeness stage, as soft and wrinkled skin fruits are ripe while rigid and smother skin fruits are less mature [6]. Various tactile sensors have been reported for fruit quality assessment [7]: a tactile sensor with piezoelectric films and strain gauges was used to evaluate and classify the surface roughness of cucumbers, cantaloupes and apples, achieving a 94% accuracy [8]. A sensor consisting of a gripper with a capacitive tactile sensor array correctly classified 88% of the mangoes according to their ripeness stage [9]. Tactile sensors inspired by the extremely sensitive mechanosensorial hair-like cilia receptors found in nature consist of an artificial cilia array inserted over a magnetic sensing element. The ciliary structures provide exquisite sensing performances due to their high aspect ratio and high surface to volume ratio [10]. Cilia sensors have been reported for multiple applications such as force and texture sensing [10]–[14] and braille reading [15]. Cilia sensors were also reported to assess fruit maturation stage, reaching a classification accuracy of 96% for apples (single cilia with 400 µm diameter and 3 mm height) and 83% for strawberries (nine cilia in a 3x3 array, each with 360 µm diameter and 1.6 mm height) [16]. Herein a texture sensor that mimics the human sense of touch is developed, capable of differentiating fruits (strawberries and blueberries) according to their skin texture, which is directly related to the fruit ripeness stage [6]. The goal was to infer fruit quality in a simple surface scan, with no data preprocessing. The sensor consists of permanent magnetic and highly elastic nanocomposite artificial cilia mounted on top of magnetic sensing elements. The nanocomposite is made of NdFeB magnetic particles incorporated into polydimethylsiloxane (PDMS), a highly elastic polymer that allows the cilia to bend and return to their original position without suffering irreversible deformations. The magnetic stray field of the cilia depends on the cilia's position and is detected by a giant-magnetoresistive (GMR) sensor.

Comparing to other techniques, this sensor presents the advantages of including magnetic sensor chips produced in large area wafers (200 mm wafer processing of GMR sensors) and soft material processing (PDMS), commonly accepted in biocompatible applications, therefore being attractive for application in industrial sorting lines.

## 2. Materials and methods

The operating principle of the device is based on GMR sensor technologies to detect the change of the cilia's magnetic stray field upon deflection. While passing the sensing system over the fruit skin the cilia bend according to its texture, thus modifying the stray field emanating from it and consequently changing the resistance of the underlying sensor.

The sensor array is made of four GMR sensing elements in a full Wheatstone bridge configuration, forming an active area of 9 mm$^2$, of which 3 mm$^2$ are the sensitive area. GMR sensors were preferred to other magnetic sensor technologies because combine higher sensitivity than AMR and lower noise than TMR technologies [17] while being fully compatible with wafer production at industrial scale.

The sensor consists of four GMR elements operating simultaneously, each composed of a series of 35 x 12 spin-valves (420 total). Each spin-valve is 40 µm long and 3 µm wide with the sensitivity along the width direction. Bonded over the sensor is the 100 cilia array, each with 150 µm in diameter and 150 µm long, spaced from each other by 100 µm.

### 2.1 GMR chip microfabrication

The GMR sensing element consists of a top-pinned spin-valve sensor [18] with the following stack (thickness in nm): Si/SiO$_2$ 100/Ta 2/NiFe 2.5/CoFe 2.8/Cu 2.8/CoFe 2.6/MnIr 7/Ta 5 deposited by Ion beam sputtering in a Nordiko 3000 tool. To obtain a linear response from the sensor, four arrays of 420 rectangular 40 x 3 µm$^2$ spin-valve sensors were defined by direct write laser lithography (DWL2.0 Heidelberg) followed by ion milling (Nordiko 3600 tool). Following, the metallic leads, consisting of 300 nm thick AlSiCu and 15 nm TiW(N$_2$) films deposited by sputtering (Nordiko 7000 tool), were patterned by laser lithography, and defined by lift-off. Finally, the chip surface was passivated with 200 nm Al$_2$O$_3$ and 200 nm SiO$_2$, deposited by magnetron sputtering, to protect against corrosion and to enable the bonding with the PDMS [19]. The sensor design is represented in figure 1.

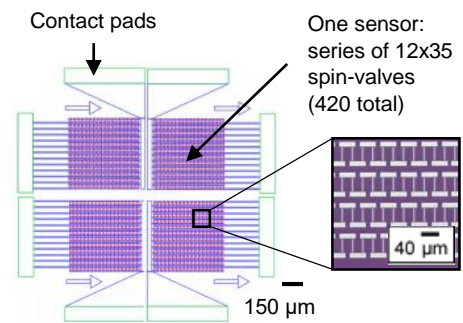

Figure 1: Top view of the microfabricated GMR sensor. The blue arrows indicate the sensitive direction of each sensor array.

2.2 Magnetic PDMS cilia fabrication

The artificial cilia array fabrication process, figure 2, started with the fabrication of a SU-8 mold, featured with a 10x10 matrix of micro-wells with 150 µm diameter and depth, using UV photolithography (UV light UH-H-254, energy intensity of 11.4 mW/cm$^2$) through a hard mask (Fig. 2a). The 150 µm thick SU-8 resist (SU-8 50, Microchem) was exposed for 52 seconds and developed with propylene glycol monomethyl ether acetate (Sigma-Aldrich) for 15.7 minutes. The mold was subjected to a hexamethyldisilazane (HDMS) silanization, by exposing it to 6 µL of HMDS on a vacuum desiccator (Bel-Art Products) for 50 minutes at room temperature (room temperature chemical vapor deposition), to decrease the SU-8-PDMS adhesion [20].

The magnetic nanocomposite was prepared by mixing NdFeB magnetic particles (Magnequench), with an average diameter of 5 µm, with PDMS (Sylgard 184, Dow Corning, Base to Curing Agent weight ratio of 15:1). A magnetic particle–to-PDMS 20% weight ratio was chosen to achieve high nanocomposite magnetization without compromising the elasticity of the cilia. The nanocomposite was casted onto the mold, removing the excess outside the micro-wells (Fig. 2b), followed by a 1-hour degassing procedure, to remove any trapped air bubbles and assist in filling the micro-wells.

A glass substrate was coated with a 1.7 µm thick poly (vinyl alcohol) (PVA) layer (20% hot water/ Mowiol 40-88, Sigma-Aldrich, weight ratio), baked at 70ºC for 5 minutes and placed in the UVO cleaner for 10 minutes, to increase the adhesion to the PDMS. A 30 µm pure PDMS (Base to Curing Agent weight ratio of 10:1) layer was spin-coated over the PVA (Fig. 2c), followed by a 45 minutes degassing procedure. The 30 µm PDMS layer enables plasma bonding to the SiO$_2$ film from the sensor. It is important that only clear PDMS is used on this layer as the presence of magnetic particles can contaminate the magnetic signal measured by the sensor, because of the background signal provided.

The SU-8 mold/nanocomposite was pressed against the glass/PVA/PDMS and cured at 70ºC for 1 hour (Fig. 2d). The glass/PVA/PDMS/nanocomposite was peeled off the mold and immersed in DI-water (in a 65ºC water bath with ultrasounds, Fig. 2e) until the PVA dissolved, releasing the cilia array (Fig. 2f). Finally, cilia with 150 µm height and diameter, were obtained "standing" on a PDMS base. To set the cilia magnetization, they were heated up to 100°C for 1 hour under a 1 T magnetic field applied in the cilia axial direction.

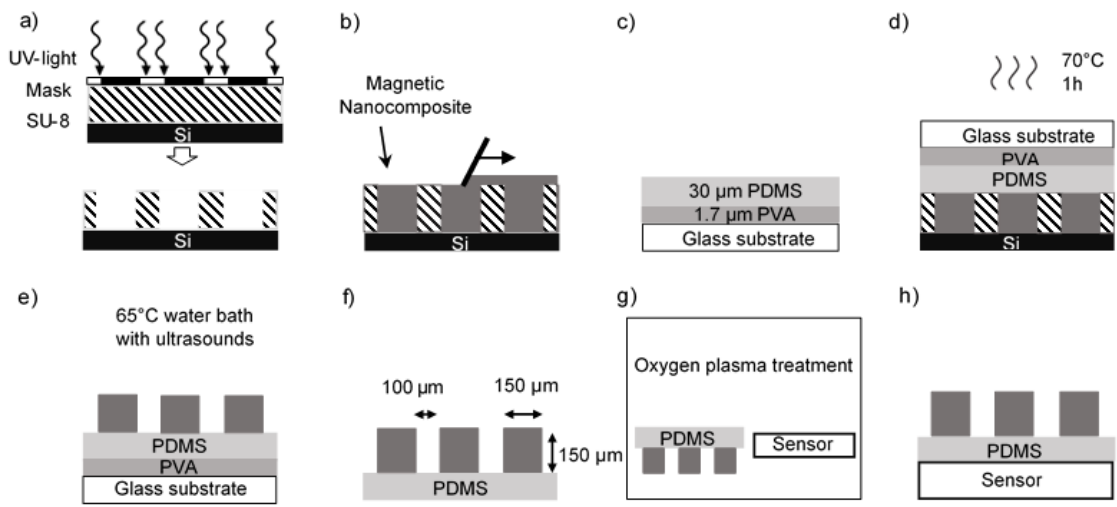

Figure 2: Schematics of the artificial cilia array fabrication process. a) SU-8 mold patterned by UV photolithography. b) Nanocomposite poured into the mold. c-d) A glass substrate was spin-coated with a PVA and a PDMS layers, pressed against the magnetic nanocomposite/mold and cured. e-f) The cilia array was peeled off from the mold and immersed in DI-water until PVA dissolved. g) Both cilia array and chip were treated by an oxygen plasma. h) The cilia array and chip were bonded. The illustration is not to scale.

2.3 Assembling and encapsulation of the chip with the cilia

Four GMR sensor arrays were individualized and assembled into a full Wheatstone bridge on a chip carrier, two of the elements had dR/dH > 0 and the two remaining ones dR/dH < 0, i.e. with antiparallel reference layers. This architecture was chosen rather than individual spin-valve sensors to mitigate temperature drifts (not affected by fruits at different temperatures, from cold storage rooms to ambient warehouse), reduce the noise and originate a bipolar output ideally centered on zero, simplifying its future integration with other electronic components [21].

Lastly, cilia array and chip were treated by an oxygen plasma (Harrick Plasma PDC-002), at medium plasma intensity for 60 seconds (Fig. 2g), and immediately brought in contact to ensure good adhesion (Fig. 2h and Fig. 3). It is important that both surfaces are thoroughly cleaned (rinsed with IPA, DI-water and blow-dried by nitrogen) before plasma treatment to improve bonding [19].

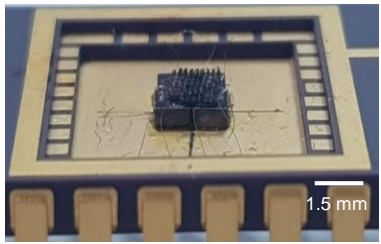

Figure 3: Optical image of the device wirebonded to a chip carrier, with the GMR sensor chip integrated with 100 artificial cilia.

2.4 Device characterization

To characterize both spin-valve array and Wheatstone bridge, a Helmholtz coil was used to apply a homogeneous magnetic field along the sensor's sensitive direction while measuring the resistance with a source meter (Keithley 2400), with a DC current bias of 1 mA. The resistance of each spin-valve was 273 Ω and of the 420 spin-valves array was 798 Ω.

3. Results and Discussion

The spin-valve sensor array showed a linear response to an applied magnetic field between -2 and 2 mT with a sensitivity of 0.07%/mT in the sensitive direction. The Full Wheatstone Bridge response to a varying magnetic field was measured (Fig. 4a), exhibiting a voltage offset of 8 mV at zero applied magnetic field (corresponding to 21.6% of the output range), due to a bridge imbalance caused by small resistance differences in between the four sensors.

The cilia sensor response to an external stimulus, with tweezers, in different directions (left, right, and front and back) is shown in figure 4b. The results (Fig. 4b) show the sensor was capable of distinguishing leftward (low voltage peak) and rightward (high voltage peak) movements and not front and backward movements, since these were executed perpendicularly to the sensor's sensitive direction. This anisotropic behavior confirms the cilia sensor directional sensitivity.

The sensor's baseline voltage differs after cilia deflection (Fig. 4b), which can be due to a collapse of some cilia upon deflection or to the detachment of the cilia array from the sensor. The first can be minimized by tuning the cilia composition, increasing the PDMS ratio and consequently cilia elasticity, and by adjusting the array design, decreasing cilia aspect ratio, and increasing spacing between them. The detachment of the cilia array from the sensor was due to irregular sensor surface onto which the array was bonded, caused by the assembly of individual components into a full Wheatstone bridge architecture, which introduces alignment errors [21]. Different oxygen plasma treatment conditions, namely increasing the plasma intensity or treatment time, could also be experimented together with a more adequate surface cleaning prior to treatment.

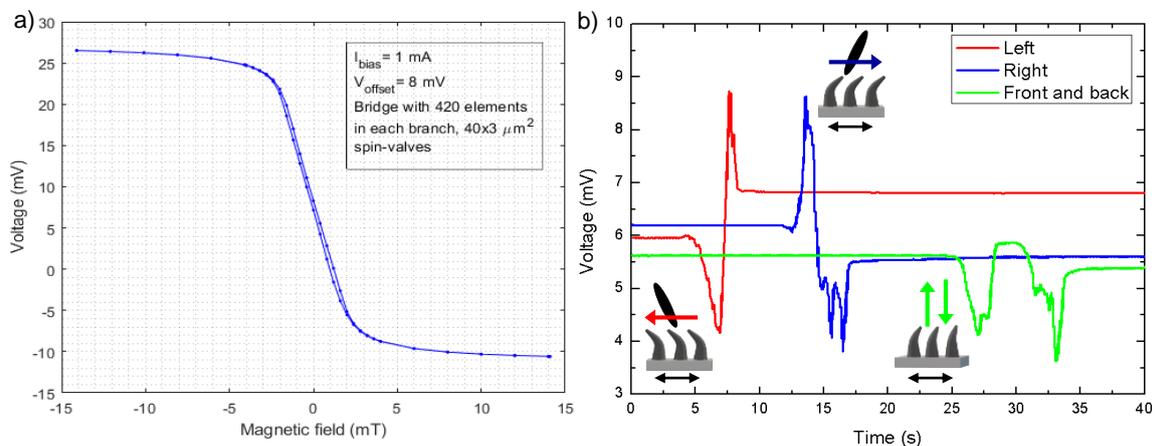

Figure 4: a) Output of the Full Wheatstone bridge to an applied magnetic field. b) Sensor response to cilia deflection. Insets: schematic of sensor stimulus motion. Black arrows indicate sensor's sensitive direction.

Lastly, a test was performed to assess the capacity of the developed sensor to detect the skin texture of fruits in different ripeness stages. Two types of fruits were analyzed, blueberries and strawberries, because their skin texture varies greatly with ripening [6] and of their small size. The fruits were separated into three categories (green, ripe and overripe) according to visual maturation assessment. The measurements were executed by fixing the sensor and manually rubbing the fruits over it, consistently at the same height (parallel to the sensitive direction). The fruits were rubbed over the sensor (first peak, Fig. 5a and 5b), slightly rotated, to measure a different location, and rubbed over the sensor again (second peak, Fig. 5a and 5b), and so on. Each fruit was scanned four times (Fig. 5).

The results (Fig. 5a and 5b) show the sensor can differentiate smooth skin fruits from wrinkled skin ones, as the maximum voltage obtained for immature fruits was 0.04 mV for blueberries and 0.05 mV for strawberries, for ripe fruits was 0.17 mV and 0.04 mV, and for overripe fruits 0.28 mV and 0.24 mV, respectively. However, for strawberries, the values obtained for green and ripe fruits are similar, making it difficult to differentiate these two maturation stages. For that, it is necessary to increase the sensor's sensitivity by altering characteristics of the cilia such as length, diameter, and Young's Modulus.

Figure 5c and 5d show the maximum voltage obtained for each fruit measurement and the average voltage verified for each maturation stage. For both fruit types averaged minimum voltage was verified for green fruits (0.14 mV for blueberries and 0.12 mV for strawberries), medium for ripe fruits (0.30 mV for blueberries and 0.23 mV for strawberries) and maximum for overripe fruits (0.58 mV for blueberries and 0.56 mV for strawberries). Accordingly, there is an increase in averaged maximum peak voltage with fruit maturation.

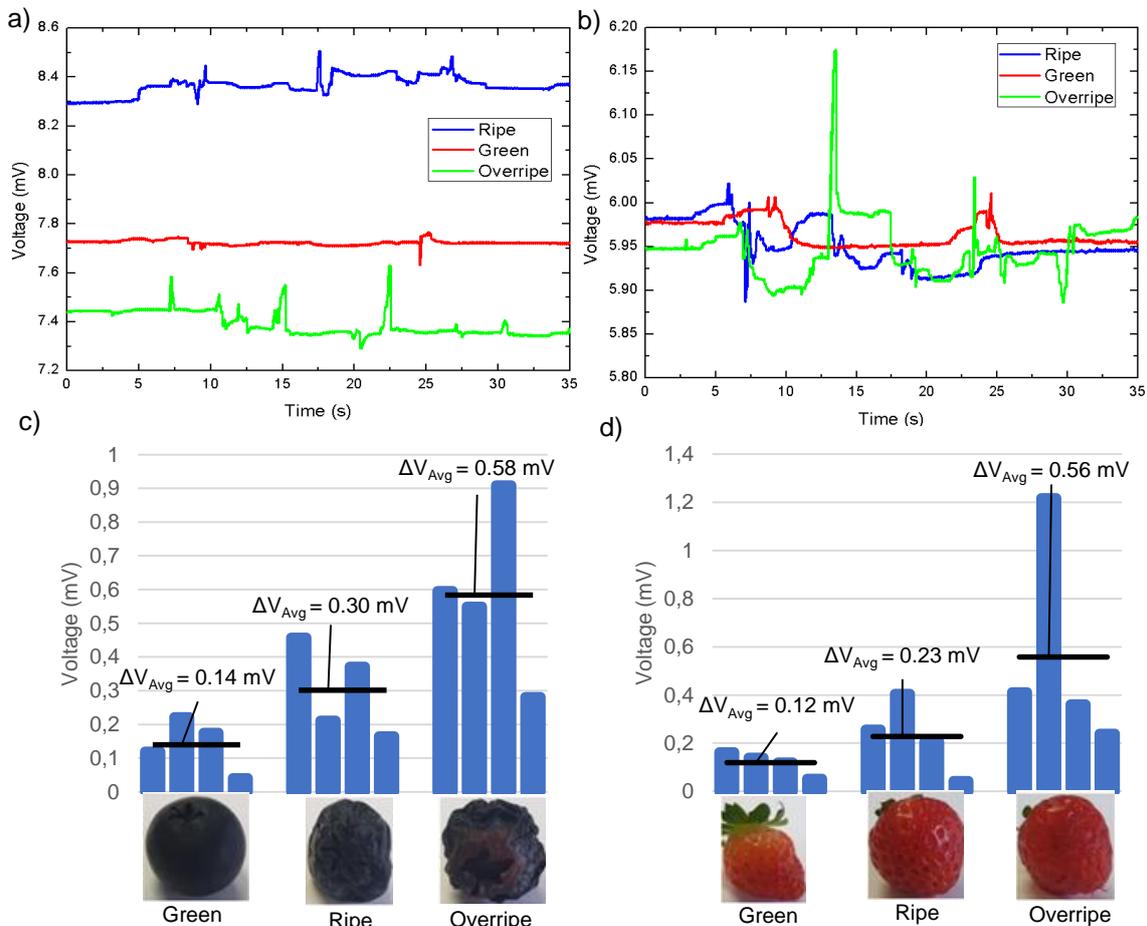

Figure 5: Skin texture measurements of a) blueberries and b) strawberries, in different ripeness stages, using the developed sensor. Each fruit was measured four times, c) and d) show the maximum voltage obtained for each measurement and the average voltage verified for each maturation stage.

## 4. Conclusion

This proof-of-concept study demonstrates that direct sensor readout could provide a fast assessment, with applications in fast quality control screening of fruits, carried out by non-specialized personnel. Spin-valve sensor arrays arranged in a Wheatstone bridge architecture detect changes of the stray field created by the permanent magnetic cilia when deflected according to the fruit skin texture.

The sensor detected the skin texture of blueberries and strawberries in different ripeness stages, showing the lowest voltage for green fruits (0.14 mV for blueberries and 0.12 mV for strawberries) and the highest voltage for wrinkled skin, overripe, fruits (0.58 mV for blueberries and 0.56 mV for strawberries). Thus, it is possible to correlate fruit ripeness stage with voltage: as fruits mature, their skin becomes more wrinkled, which when rubbing it over the sensor causes a more accentuated cilia deflection, resulting in a higher voltage.

An advantage of the cilia is their permanent magnetic behavior which facilitates future sensor integration since it does not require an external magnetic field bias. Future work includes improving the sensor architecture, by increasing cilia array and sensing area dimensions, and sensitivity to be more tailored to fruit quality assessment. Also, fixing the fruit in a moving stage to remove the operator interference and reduce output signal noise.

The developed sensor may, in the future, be integrated into a robot's fingertips for it to evaluate the skin texture of fruits and grade them according to the ripeness stage. This way, fruits in a more advanced maturation stage will not be packed together with less ripe fruits, which will decrease their premature ripening and consequently improve product durability, quality, and value.

## Acknowledgments

This work was partially supported by Fundacao para a Ciencia e Tecnologia through Project MagScopy4IHC-LISBOA-01-0145-FEDER-031200. P. Ribeiro acknowledges FCT for his PhD grant SFRH/BD/130384/2017. INESC-MN acknowledges FCT funding through National Infrastructure Roadmap Unit (UID/05367/2020) through plurianual BASE and PROGRAMATICO financing.